\documentclass[reprint,
superscriptaddress,
preprintnumbers,
amsmath,amssymb,
aps,
prb,
floatfix,
]{revtex4-2}
\usepackage{xcolor}
\usepackage{graphicx}
\usepackage{dcolumn}
\usepackage{bm}
\usepackage{hyperref}
\usepackage[mathlines]{lineno}
\hypersetup{colorlinks=true, linkcolor=blue, filecolor=blue, urlcolor=blue, citecolor=blue }
\newcommand{\etal}{{\textit{et al. }}}

\begin{document}

\title{Atomistic insights into the mixed-alkali effect in phosphosilicate glasses}

\author{Achraf Atila}
\email{achraf.atila@fau.de;\\ achraf.atila@gmail.com}
\affiliation{Department of Materials Science $\&$ Engineering, Institute I: General Materials Properties, Friedrich-Alexander-Universit\"{a}t Erlangen-N\"{u}rnberg, 91058 Erlangen, Germany}
\affiliation{Computational Materials Design, Max-Planck-Institut f\"{u}r Eisenforschung Max-Planck-Str. 1, 40237 D\"{u}sseldorf, Germany}
\author{Youssef Ouldhnini}
\affiliation{LS2ME, Facult\'{e} Polydisciplinaire Khouribga, Sultan Moulay Slimane University of Beni Mellal, B.P 145, 25000 Khouribga, Morocco}
\author{Said Ouaskit}
\affiliation{Laboratoire de Physique de la Mati\`{e}re Condens\'{e}e, Facult\'{e} des Sciences Ben M'sik, University Hassan II of Casablanca, B.P 7955, Av Driss El Harti, Sidi Othmane, Casablanca, Maroc}
\author{Abdellatif Hasnaoui}
\affiliation{LS2ME, Facult\'{e} Polydisciplinaire Khouribga, Sultan Moulay Slimane University of Beni Mellal, B.P 145, 25000 Khouribga, Morocco}

\date{\today}

\begin{abstract}
Oxide glasses have proven useful as bioactive materials, owing to their fast degradation kinetics and tunable properties. Hence, in recent years tailoring the properties of bioactive glasses through compositional design have become the subject of widespread interest for their use in medical application, e.g., tissue regeneration. Understanding the mixed alkali effect (MAE) in oxide glasses is of fundamental importance for tailoring the glass compositions to control the mobility of ions and, therefore, the glass properties that depend on it, such as ion release, glass transition temperature, and ionic conductivity. However, most of the previously designed bioactive glasses were based on trial-and-error, which is due to the complex glass structure that is non-trivial to analyse and, thus, the lack of a clear picture of the glass structure at short- and medium-range order. Accordingly, we use molecular dynamics simulations to study whether using the MAE can control the bioactivity and properties of 45S5 glass and its structural origins. We showed that the network connectivity, a structural parameter often used to access the bioactivity of silicate glasses, does not change with Na substitution with Li or K.
On the contrary, the elastic moduli showed a strong dependence on the type of the modifier as they increased with increasing mean field strength. Similarly, the mobility of the glass elements was significantly affected by the type of modifier used to substitute Na. The change of the properties is further discussed and explained using changes at the short- and medium-range structure by giving evidence of previous experimental findings. Finally, we highlight the origin of the non-existence of the MAE, the effect of the modifier on the bioactivity of the glasses, the importance of dynamical descriptors in predicting the bioactivity of oxide glasses, and we provide the necessary insights, at the atomic scale, needed for further development of bioactive glasses.

\end{abstract}

\maketitle

\section{Introduction}
Oxide glasses have attracted much interest owing to their unique, highly tailorable properties \cite{musgraves2019springer}. In addition, a large number of compositions in the form of a combination of all elements from the periodic table enable control of oxide glasses properties \cite{Shakhgildyan2020} and, in turn, allow for the development of new glasses for many technological and medical applications \cite{Shakhgildyan2020, musgraves2019springer}.

Since the development of the first bioactive glasses (45S5 bioactive glasses or Hench glasses) by Hench and his colleagues in the late sixties \cite{Hench1971}, much research has been devoted to developing new oxide glass with good bioactivity. The 45S5 bioactive glass has been used in clinical applications starting from the mid-1980s \cite{Hench2004}. The 45S5 glass gained its good bioactivity from its ability to form hydroxycarbonate apatite (HCA) as a surface layer enabling the bonding to the bones, releasing ions, and more importantly, degrade in the body \cite{Baino2018}. When implanted in the human body, a process of ion exchange between the glass surface and the surrounding biological fluid starts, leading to the formation of a bone-like HCA layer, which is replaced partially by the bone after long-term implantation \cite{Baino2018, Hench2004}. 
Even though their excellent bioactive properties, the Hench type bioactive glasses suffer from low mechanical strength and fracture toughness, restricting their usage to only applications that are not exposed to high loads \cite{Ouldhnini2021a}. Furthermore, many developed bioactive glasses resulted from empirical studies, which hinders the further development of oxide glasses as bioactive materials. Nevertheless, new bioactive glasses with enhanced mechanical properties were developed using different routes, such as producing composites by combining oxide glasses with polymeric phases \cite{Karan2021} or processing techniques like cooling under pressure \cite{Ouldhnini2021a}. 

Controlling the bioactivity through the control of the ion release is of great importance, and it has been the focus of many papers. For instance, Br\"{u}ckner \textit{et al.} investigated the effect of modifier size on the ion release and the apatite formation in 45S5 bioactive glass. They partially replaced the Na$_2$O with either Li$_2$O or K$_2$O, and as hypothesized, the network connectivity of the glass is expected to be the same. In contrast, the size of the added cations (Li or K) affected the bioactivity of this glass. Moreover, the change in the glass properties with partial replacement of one alkali by another is known as the MAE \cite{Swenson2003, Yu2017, Wang2017, Onodera2019, Lodesani2020, Crovace2021}, and it has been shown to affect the ion release of silicate glasses where the mixed alkali silicate compositions show lower ion release compared to the glasses having only one alkali element \cite{DILMORE1978}.
The addition of different modifiers to the original 45S5 composition was heavily investigated to tweak and enhance the properties of this bioactive glass. Arepalli \textit{et al.} investigated the effect of low barium (BaO) content on the bioactivity, thermal, and mechanical properties of 45S5 bioactive glasses \cite{Arepalli2015}. They showed that the addition of BaO leads to a decrease in crystallization temperature, increasing the flexural strength, and improved hemolysis compared to the pristine (unmodified) 45S5 \cite{Arepalli2015}. Another study showed that the 45S5 glass containing 1.35 mol\% of BaO have improved anti-inflammatory properties compared to the pristine 45S5 \cite{Majumdar2021}. Karan \textit{et al.} \cite{Karan2021} investigated the effect of lithium oxide substitution on the structure, in-vitro chemical dissolution, and mechanical properties of 45S5 based glasses and glass-ceramics\cite{Karan2021}. They found that the mechanical properties increased with increasing Li$_2$O content in the glass and glass-ceramics, which was attributed to compact glass structure in the presence of small-sized Li$^+$ ions \cite{Karan2021}. In-vitro study of glass-ceramics samples in simulated body fluid solution showed that carbonated hydroxyapatite was formed on the glass-ceramics surface.

Although experiments provided great details on the bioactivity of the glasses, the atomic-scale picture of the events that led to the bioactivity is not fully clear. Molecular dynamics (MD) simulations provide quantitative and qualitative insights at the atomic scale that can be used to understand the composition--processing--structure--properties relationship in glasses. In this regard, the effect of processing on structure, elasticity, and diffusion behaviour of 45S5 bioactive glass has been studied using MD simulations \cite{Ouldhnini2021a, Ouldhnini2021b}. Furthermore, the effect of alkaline earth oxides, especially Sr, on the structure, diffusion, and bioactivity of 45S5 was studied using MD simulations \cite{Xiang2011, Du2012, Du2016, Xiang2013}. These studies showed that partial substitution of CaO by SrO have a beneficial effect on tissue growth and enhances bioactivity by increasing the dissolution rate. This is directly linked to a weakening of the glass network as SrO have a field strength lower than CaO and thus enables diffusion pathways with lower energy barriers for cation/water exchange\cite{Xiang2011, Du2012, Du2016}.
Moreover, there is always a need to tailor, enhance, or design oxide glasses that have a combination of good mechanical properties and bioactivity. In this context, we study the effect of Li$_2$O/Na$_2$O and K$_2$O/Na$_2$O substitution on the elastic, dynamic, and structural properties of 45S5 bioactive glasses at the atomic scale, and we provide an atomistic understanding that is still missing to the observed changes. Finally, the results are compared with the data available in the literature. 

The remainder of this paper is organized as follows: In Sec. \ref{methods}, we describe the procedure followed to obtain the results. The calculated properties are presented in Sec. \ref{results}. In Sec. \Ref{discussion} we discuss the results, and we suggest possible explanations for the obtained results. Concluding remarks are given in Sec. \ref{conclusion}.

\section{\label{methods}Methods}
\subsection{Interatomic potential model}
The Pedone \etal potential \cite{Pedone2006} was used to define the interactions between atoms. In this potential, the particles are treated as charge points interacting via Coulomb forces, a Morse function to describe the short-range interactions between pairs of atoms, and an additional $r^{-12}$ repulsive contribution necessary to model the interactions at high pressure and temperature. This potential gives a realistic agreement with available experimental data as mentioned in the literature \cite{Pedone2006, Ghardi2019, Atila2019b, Atila2020c, Atila2020a, Ouldhnini2021a, Ouldhnini2021b} as it was designed to reproduce structural and mechanical properties of a wide range of oxide glasses. Potential parameters and partial charges are given in the Ref. \cite{Pedone2006}. An interaction cutoff of 5.5 $\AA$ was used for short-range interactions, while  Coulomb interactions are calculated by adopting Fennell damped shifted force (DSF) model \cite{Fennell2006} with a damping parameter of 0.25 $\AA^{-1}$ and of 8.0 $\AA$ as a long-range cutoff. 
\subsection{Glass preparation}
We simulated a series of nine mixed alkali 45S5 glasses (CaO)$_{26.9}$--(Na$_2$O)$_{(24.4-x)}$--(M$_2$O)$_{x}$--(SiO$_2$)$_{46.1}$--(P$_2$O$_5$)$_{2.6}$ (with M = Li or K, and x = 0, 6.1, 12.2, 18.3, and 24.4 mol\%) with classical MD, using the LAMMPS package \cite{Plimpton1995}. These compositions mimics the ones considered in the experimental work of Br{\"u}ckner \textit{et al.}\cite{Brckner2016}.
All samples consisting of around 10000 atoms placed randomly (with no unrealistic overlap) in a cubic simulation box under periodic boundary conditions with the box edges were chosen to reproduce the experimental glass density as given in Ref. \cite{Tylkowski2013}. The samples were equilibrated at a high temperature (4000 K) for 500 ps to get an equilibrated melt. The melts were subsequently quenched to 300 K using a cooling rate of 1 K/ps. After quenching, the glass was further equilibrated at 300 K in the NPT ensemble (isothermal-isobaric ensemble) for 1 ns and a 100 ps in NVT ensemble (canonical ensemble) for statistical averaging. The results presented in this paper are averaged over 100 configurations (from the last 100 ps of the NVT run) unless otherwise mentioned. Comparison between the room temperature experimental glass density and the one obtained from our simulations after the NPT relaxation is given in supplementary material \cite{SuppMat} Tab. S1. The temperature and pressure were controlled using the Nosé-Hoover thermostat and barostat \cite{Hoover1985, Parrinello1981}. The equations of motion were solved using the velocity-Verlet algorithm with a timestep of 1 fs.

\subsection{Mean squared displacement and diffusion coefficient}
The dynamic of the samples was calculated at a temperature higher than the glass transition temperatures by computing the mean-squared displacement (MSD),
\begin{equation}
    MSD = \left< |r(t) - r(0)|^2 \right>
\end{equation}
where $r(0)$ is the initial position at time $t$ = 0 and $r(t)$ denotes the position at a time t, at temperatures ranging between 1500 K to 2200 K with a resolution of 100 K. The MSD was calculated using trajectories obtained during NVT runs for 2 ns and using a time step of 1 fs at each temperature. At the beginning of each simulation, the sample was equilibrated in the NPT ensemble at the desired temperature and zero MPa for 100 ps. The diffusion coefficient D was obtained using Einstein's equation,
\begin{equation}
    D = \lim_{t \to +\infty} \frac{MSD}{6t}
\end{equation}
and it was averaged over the last 200 ps of each run. The length of the simulation time used in this work is long enough for the MSD to be in the linear regime and for the diffusion coefficient to converge (see Fig.~S1 in the supplementary material \cite{SuppMat}).

\subsection{Calculation of elastic moduli}
To calculate the elastic moduli, we used the second derivative method \cite{Atila2019b}. This method enables us to compute the stiffness matrix as well as the compliance matrix. Using a single-point energy calculation, the stiffness matrix elements are obtained by:
\begin{equation}
\label{eq:Cij}
C_{ij} = \frac{1}{V}\frac{\partial^{2}U}{\partial\varepsilon_{\alpha}\partial\varepsilon_{\beta}}
\end{equation}
For cubic isotropic materials, there are only two independent parameters of the stiffness matrix ($C_{11}$ and $C_{44}$) \cite{Pedone2006, Atila2019b} and $C_{12} = C_{11} - 2C_{44}$. Following the Voigt convention, the bulk, shear and Young moduli are give by Eqs. \ref{eq:Bvoigt} -- \ref{eq:young}, 
\begin{equation}\label{eq:Bvoigt}
B_{Voigt} = \frac{C_{11}+2C_{12}}{3}
\end{equation}
 
\begin{equation}\label{eq:Gvoigt}
G_{Voigt} = C_{44}
\end{equation}
 
\begin{equation}\label{eq:young}
E = \frac{9BG}{3B+G}
\end{equation}

The computed elastic moduli reported in this paper were calculated using molecular statics through energy minimization. The obtained glass structures at 300 K were subjected to an energy minimization to the closest energy minimum in the potential energy landscape using the conjugate gradient algorithm. The minimized structures were deformed in each of the six directions in both positive and negative directions, and the stress tensor was measured. This operation was repeated 100 times using different uncorrelated initial configurations to get an averaged elastic moduli.

\subsection{Structural descriptor F$_{net}$}
In previous studies, many authors used the F$_{net}$, which is a structural descriptor for the estimation of the glass strength. This parameter, which was originally proposed by Lusvard \textit{et al.} \cite{Lusvardi2009}, takes into account both structural and energetic properties of the glass. Recently, a modified version of the F$_{net}$ was proposed \cite{Du2021}. In this new definition (Eq. \ref{FnetBS}), the energetic properties of the glass are the single bond strength (SBS) as calculated by Sun \cite{Sun1947} instead of the bond enthalpy of diatomic molecules used in the original formulation. 

\begin{equation}
\label{FnetBS}
F_{net} = \frac{1}{N} \sum^{cations}_X n_X.CN_{XO}.SBS_{XO}.M_{NC}
\end{equation}

Where $N$ is the total number of atoms, $n_X$ is the number of cations of type $X$, $CN_{XO}$ is the average coordination number of the pair $X$--$O$, and $M_{NC}$ is the overall network connectivity that is $NC = NC_{Si} + NC_{P} = \sum^4_{n=0} n\times Q^n_{Si/P}$, and Q$^n$ represent the fraction of n bridging oxygen per glass former tetrahedron which gives more accurate evaluation. This modified version of the F$_{net}$ was shown to have an excellent correlation with physical properties and dissolution rate of many glasses \cite{Du2021, Lu2019, Lu2020}.

\section{\label{results} Results}
\subsection{Diffusion}
\begin{figure*}[htb!]
\includegraphics[width=\textwidth]{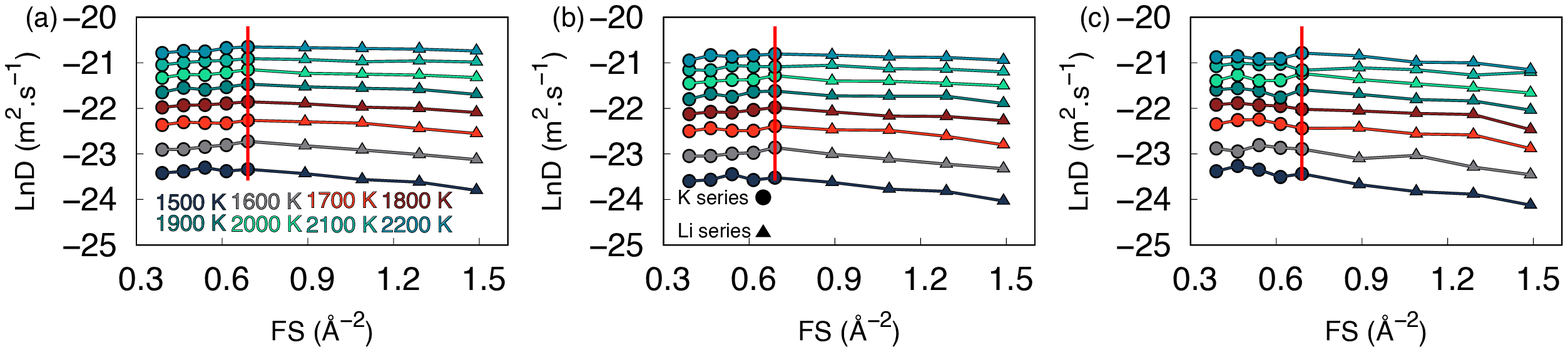}
\caption{\label{D_OSiP}The diffusion coefficient of (a) O, (b) Si, and (c) P as a function of the mean-field strength and for different temperatures in the mixed-alkali 45S5 bioactive glass. Filled circles indicate data for potassium mixed-alkali 45S5, while filled triangles are for lithium mixed-alkali 45S5 bioactive glasses. The red vertical lines indicate the pristine 45S5 bioactive glass. The diffusion data is averaged over the last 200 ps from an NVT run at each temperature, and the error bars are within the size of the symbol and are removed for clarity.}
\end{figure*}
The self-diffusion coefficients of all elements of the glasses studied here are determined from the mean squared displacement (MSD) (See Supp Mat Fig.~S1 for the pristine 45S5 composition) that was measured from MD simulated trajectories at temperatures between 1500 -- 2200 K with a resolution of 100 K and for a duration of 2 ns. The MSD curves displayed in the supplementary material \cite{SuppMat} Fig.~S1 show that the simulation time was long enough for the atoms to diffuse and get a linear regime. Moreover, the MSDs of O, Si, and P showed that the glass former matrix diffused an order of magnitude less than the modifiers, namely Li, Na, K, and Ca. In addition to that, it is clear that the addition of Li$_2$O helped reduce the diffusivity of the glass network former atoms, while we observed an increase in the case of K$_2$O addition.

\begin{figure*}[htb!]
\includegraphics[width=\textwidth]{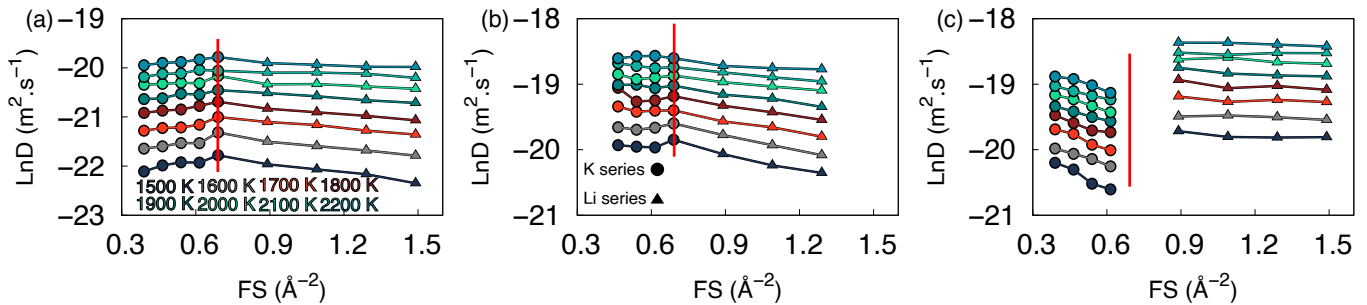}
\caption{\label{D_LiNaKCa}The diffusion coefficient of (a) Ca, (b) Na, and (c) Li or K as a function of the mean-field strength and for different temperatures in the mixed-alkali 45S5 bioactive glass. Filled circles indicate data for potassium mixed-alkali 45S5, while filled triangles are for lithium mixed-alkali 45S5 bioactive glasses. The red vertical lines indicate the pristine 45S5 bioactive glass. The diffusion data is averaged over the last 200 ps from an NVT run at each temperature, and the error bars are within the size of the symbol and are removed for clarity.}
\end{figure*}

The bioactivity of oxide glasses has been shown to be related to the mobility of ions within the glass, and their ability to migrate to the glass surface and dissolve in the body \cite{Brckner2016, Tilocca2017}, leading to a necessity of checking the in vitro cytocompatibility of the glass \cite{Brckner2016}. The diffusion coefficients of all elements are plotted in Fig.~\ref{D_OSiP} and Fig.~\ref{D_LiNaKCa} as a function of the mean-field strength, which was calculated as ((1-x)FS$_{Na-O}$ + xFS$_{M-O}$), where $M$ is Li or K, and x is the fraction of modifier used to substitute N$_2$O. As shown by the plots in Fig.~\ref{D_OSiP} and Fig.~\ref{D_LiNaKCa} the self-diffusion coefficients of all elements depend essentially on the composition and temperature. The self-diffusion coefficient of the glass network former O, Si, and P, increases with increasing temperature, which is the expected behaviour. In the K-45S5 series, the diffusion of those elements stayed almost constant with the increase of K$_2$O content while it showed a slight decrease in the Li-45S5 series with the increase of Li$_2$O content (increasing mean field strength).
On the other side, the self-diffusion coefficient of the modifiers does not follow the same trend. In Fig.~\ref{D_LiNaKCa}(a), we show that the mobility of Ca ions increases with decreasing K$_2$O content with a maximum in the pristine 45S5, and it decreases with increasing Li$_2$O content. In contrast to Ca, Na ions show almost no change in diffusion coefficient in the K-45S5 series and become less mobile with the increase of the Li$_2$O content as indicated by the decrease of the diffusion coefficient shown in Fig.~\ref{D_LiNaKCa}(b). For the diffusion coefficient of Li and K shown in Fig.~\ref{D_LiNaKCa}(c), D$_K$ increase with increasing K$_2$O and D$_{Li}$ showed no dependence on Li$_2$O content.

\begin{figure}[htb!]
\includegraphics[width=\columnwidth]{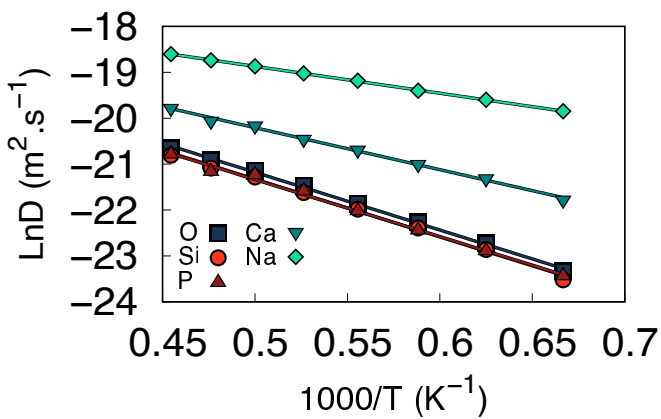}
\caption{\label{Arrhenius_45S5}Diffusion coefficients of O, Si, P, Na, and Ca as a function of temperature for the studied 45S5 bioactive glass. The symbols represent the simulated data, and the lines are fit to the Arrhenius law. The diffusion data is averaged over the last 200 ps from an NVT run at each temperature, and the error bars are within the size of the symbol and are removed for clarity.}
\end{figure}

As mentioned previously, the self-diffusion coefficients of each element were quantitatively calculated from the slope of the MSD versus time in the linear regime and in all temperature range. The lnD$_X$, where X is one of the glass elements, were plotted vs 1000/T, and it follows an Arrhenius behaviour that is depicted by the following equation,
\begin{equation}
lnD = lnD_0 - \frac{\Delta E_a}{k_BT}
\label{Arrheniuseq}
\end{equation}
where E$_a$ is the activation energy, D$_0$ is the pre-exponential factor, k$_B$ is the Boltzmann constant, and T is temperature \cite{Atila2020b}. We can see that the activation energy is the slope of the linear fitting between lnD and 1000/T; this follows perfectly the Arrhenius law (eq.~\ref{Arrheniuseq}) as is shown in Fig.~\ref{Arrhenius_45S5} for all elements in the pristine 45S5. The calculated values of the correlation factor R$^2$ in all glasses and for all elements were higher than 0.98, showing excellent linearity that validates the calculation of the activation energies for self-diffusion.

\begin{figure*}[htb!]
\includegraphics[width=\textwidth]{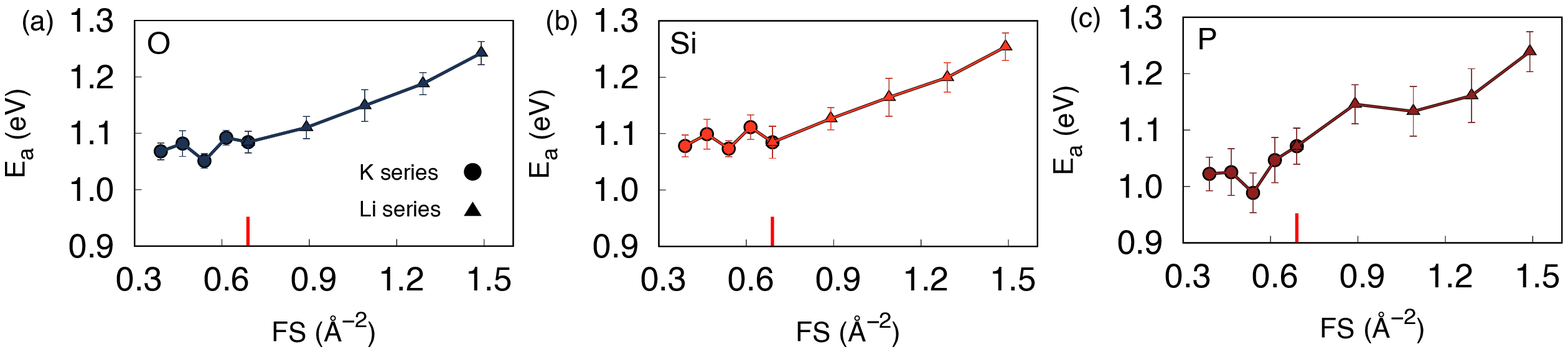}
\caption{\label{Ea_OSiP}The activation energies (E$_a$) of (a) O, (b) Si, and (c) P as a function of the mean-field strength in the mixed-alkali 45S5 bioactive glass. The E$_a$ was calculated from the fitting of the lnD vs 1000/T with the Arrhenius function. Filled circles indicate data for potassium mixed-alkali 45S5, while filled triangles are for lithium mixed-alkali 45S5 bioactive glasses. The red vertical lines indicate the pristine 45S5 bioactive glass. The error is estimated from the fit.}
\end{figure*}

\begin{figure*}[htb!]
\includegraphics[width=\textwidth]{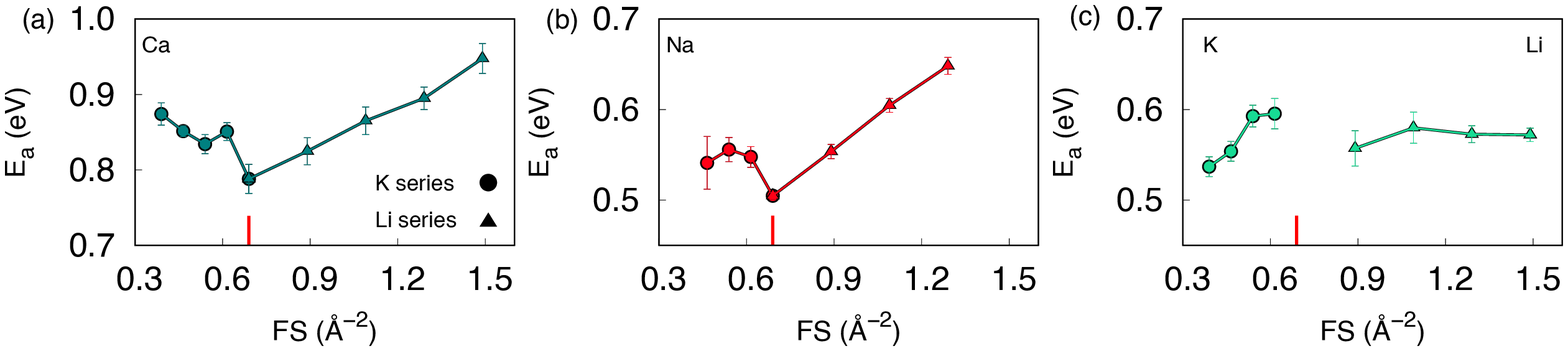}
\caption{\label{Ea_CaNaLiK}The activation energies (E$_a$) of (a) Ca, (b) Na, and (c) Li or K as a function of the mean-field strength in the mixed-alkali 45S5 bioactive glass. The E$_a$ was calculated from the fitting of the lnD vs 1000/T with the Arrhenius function. Filled circles indicate data for potassium mixed-alkali 45S5, while filled triangles are for lithium mixed-alkali 45S5 bioactive glasses. The red vertical lines indicate the pristine 45S5 bioactive glass. The error is estimated from the fit.}
\end{figure*}

The activation energies for self-diffusion for O, Si, and P (Fig.~\ref{Ea_OSiP}) are slightly affected by the content of K$_2$O while they increase significantly with increasing Li$_2$O content. The behaviour of the activation energies of the modifiers does strongly depend on the type of modifier used to substitute Na, as shown in Fig.~\ref{Ea_CaNaLiK}. Generally speaking, the activation energies are lower than those obtained for O, Si, and P atoms, as expected. The substitution of Na$_2$O by K$_2$O or Li$_2$O increased the activation energy for Ca and Na self-diffusion with the minimum found in the pristine 45S5 glass. We observe a clear non-linearity of the behaviour of the activation energy of Na migration with mean FS in the K-45S5 glass, highlighting the presence of the MAE, while it is almost not present in the case of Li-45S5. The activation energy of Li self-diffusion is slightly affected by the Li$_2$O content, and that of K decreases significantly with increasing K$_2$O content. 

\subsection{Elastic properties}
\begin{figure}[htb!]
\includegraphics[width=\columnwidth]{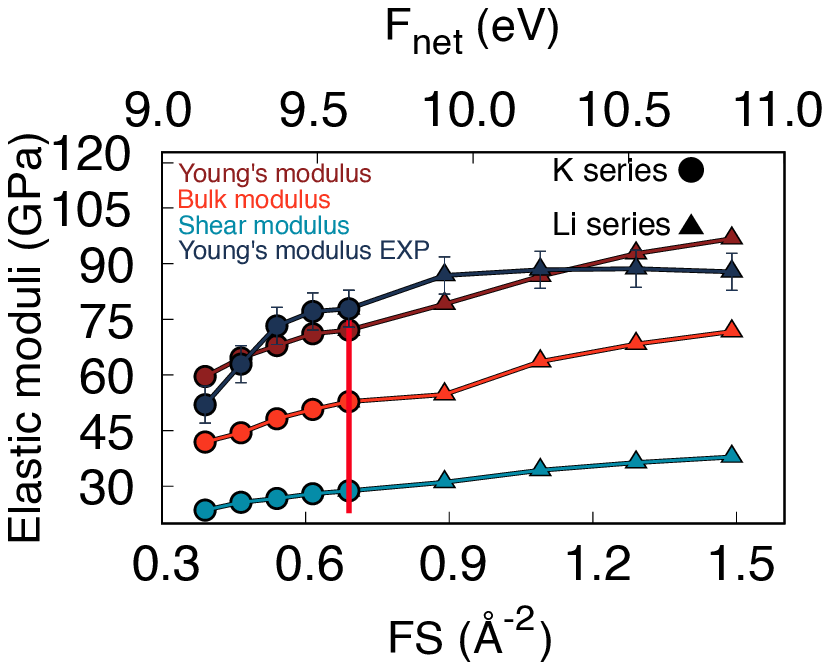}
\caption{\label{Elastic}The elastic moduli as a function of the mean-field strength in the mixed-alkali 45S5 bioactive glasses. Experimental values of the Young's modulus are taken from Ref. \cite{Tylkowski2013}.  Filled circles indicate data for potassium mixed-alkali 45S5, while filled triangles are for lithium mixed-alkali 45S5 bioactive glasses. The red vertical lines indicate the pristine 45S5 bioactive glass. The elastic moduli are averaged over 100 initial configurations, the maximum error obtained for the elastic moduli was 0.5 GPa, and the error bars are not shown.}
\end{figure}

Figure~\ref{Elastic} shows the elastic moduli (Young's modulus, bulk modulus, and shear modulus) as a function of the F$_{net}$ and mean field strength (FS). The calculated elastic moduli are in good agreement with available experimental data \cite{Tylkowski2013, Karan2021}. The elastic moduli decrease with increasing K$_2$O content and increase with increasing Li$_2$O content, highlighted by the increase of the elastic moduli with increasing mean FS. This can be seen as an indication of the increasing bond strength and elasticity of these glasses. Moreover, the rate of the increase of the elastic moduli with decreasing K$_2$O content in the K-45S5 series is slightly higher than the rate of increase of the elastic moduli with increasing L$_2$O content in the Li-45S5 series. Thus we can tweak the elasticity of the 45S5 bioactive glass by entirely or partially replacing Na with Li or K. We observed a similar trend of the elastic moduli with the F$_{net}$, and in good accordance with previous studies that showed that the elastic moduli are positively correlated with F$_{net}$ \cite{Du2021}. 

\begin{figure*}[htb!]
\includegraphics[width=\textwidth]{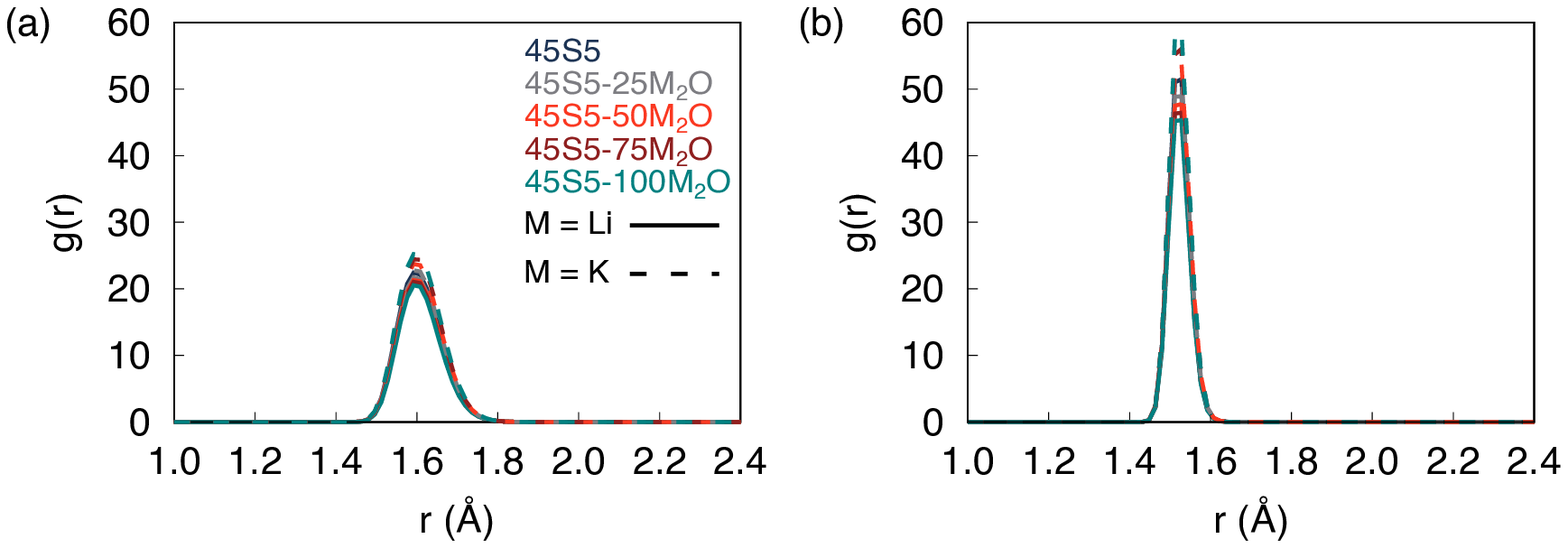}
\caption{\label{RDFSiPO}Si--O (solid lines) and P--O (dashed lines) pair distribution functions in both (a) Li mixed-alkali and (b) K mixed-alkali 45S5 bioactive glasses at 300 K.}
\end{figure*}
\subsection{Structure}
\subsubsection{Local structure of the glass former atoms}
The local structure of the glass former atoms (Si and P) in the glasses is analyzed using the pair distribution functions as depicted in Fig.~\ref{RDFSiPO} where we focus only on the first peak. The average bond distances Si--O (1.6 $\AA$ ) and P--O (1.52 $\AA$ ) did not show any dependence of the content or type of the modifier used to substitute Na, which is in accordance with the data reported in the literature \cite{Atila2019b}. However, the broadness of the Si--O peak decreased as a function of the mean field strength, which means that with increasing Li$_2$O content, the Si--O bond lengths are more homogeneous while increasing K$_2$O they are more heterogeneous. The integration of the pair distribution functions to a specific cutoff, defined as the first minimum of each RDF, enable us to obtain the mean coordination numbers of the network former atoms in the first coordination shell, that means how many oxygen atoms surround Si or P. At the same time, its evolution with distance is called cumulative coordination number. This minimum is found to be 2.0 $\AA$ for Si--O and 1.8 $\AA$ for P--O. The mean coordination number of Si--O and P--O shows a plateau indicating very well-defined first coordination shell. Both Si and P are surrounded by four oxygen atoms and it is independent of the composition in the studied glasses. This is in good accordance with previous results \cite{Ouldhnini2021a}.

The Si--Si, P--P, and Si--P PDFs generally present the distance of separation between the center of SiO$_4$-SiO$_4$, PO$_4$--PO$_4$, and SiO$_4$-PO$_4$ tetrahedra. These PDFs show peaks around 3.16 $\AA$ for both Si--Si and Si--P pairs, while for P--P, the PDFs are not clear, which is due to the low P--P content (see supplementary material \cite{SuppMat} Fig.~S2). The corresponding cumulative coordination numbers of Si--Si and Si--P have values around 2.0 and 0.1 as shown in the supplemental material Fig.~S2. For the P--P pair, the distribution does not show any correlations because PO$_4$ are distributed in the network as isolated units. In addition, the PDFs, the bond angle distributions (BADs) provide details on the inter and intra polyhedral angles and are shown in the supplementary material \cite{SuppMat} Fig.~S3. The O--P--O BAD is centred around 109.4\textdegree\  and show no dependence on the composition in the composition range studied here. The O--Si--O and O--P--O BADs show the distribution of the angles inside SiO$_4$ and PO$_4$ tetrahedra and are shown in Fig.~S3. The O--Si--O distribution is centred around 109\textdegree\  and have a small shift of 1\textdegree\  toward lower angles in the Li45S5 while it is unaffected in the K-45S5. The Si--O--Si BADs have a broad distribution centred around an angle of 146\textdegree\ . For the Si--O--P BADs is affected by the content of Li or K. In the pristine 45S5, it is a broad distribution centred around 160\textdegree\  and shifts toward small angles with the increase of the Li or K content; moreover, the distribution shows a double peak in the glasses when Li or K. fully substitutes Na We should stress that P--O--P linkage does not exist in the 45S5 glasses, which is in accordance with the data available in the literature \cite{Ouldhnini2021a, Bhaskar2020}.

\begin{figure}[htb!]
\includegraphics[width=\columnwidth]{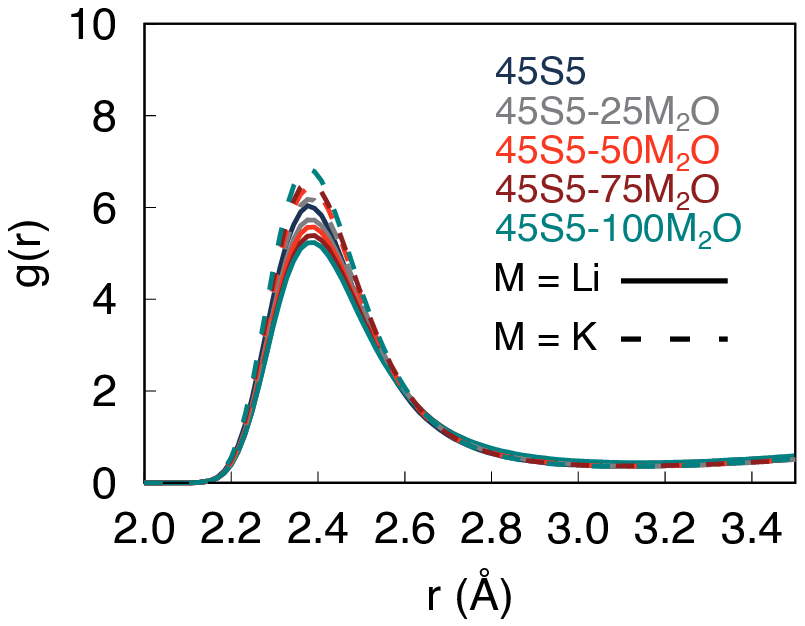}
\caption{\label{RDFCaO}Ca--O pair distribution functions in both Li mixed-alkali (solid lines) and K mixed-alkali (dashed lines) 45S5 bioactive glasses at 300 K.}
\end{figure}

\begin{figure*}[htb!]
\includegraphics[width=\textwidth]{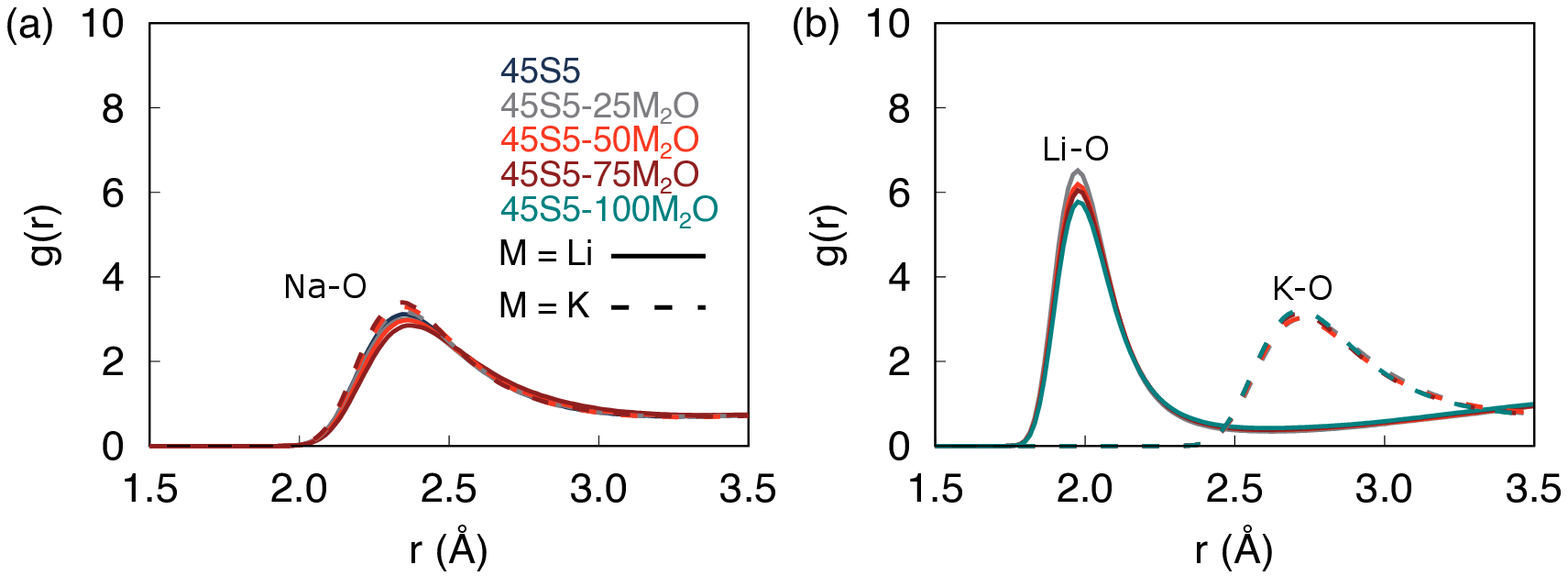}
\caption{\label{RDFLiNaK}(a) Na--O and (b) Li and K--O pair distribution functions in both Li mixed-alkali (solid lines) and K mixed-alkali (dashed lines) 45S5 bioactive glasses at 300 K.}
\end{figure*}

\subsubsection{Local structure of the glass modifiers}
Figure.~\ref{RDFCaO} show the change of the Ca--O RDF with the type and content of the modifier used to substitute Na. With increasing K$_2$O content, the Ca--O RDF becomes broader, while with increasing Li$_2$O content, it becomes more narrowed compared to the pristine 45S5 glass. The mean coordination number of Ca--O pair, shown in Tab.~S2, increases with increasing mean field strength, from 5.97 in the glass with Na fully substituted by K to 6.77 in the glass with Na entirely substituted with Li. The Na--O behaves similarly to the Ca--O pair for the RDF and shows increasing mean coordination with increasing mean FS. The Li--O and K--O RDFs showed less dependence on the content of Li or K, while the Li--O mean coordination number increased with increasing Li$_2$O content and that of K--O decreased with increasing K$_2$O content. 
If we look at the BAD of O-X-O and X-O-X, where X stands for Li, Na, K, or Ca, as plotted in Figs.~S3 and Fig.~S4, we can see that the environment of the modifiers is more complex. This is due to the high coordination states that can be attained by these cations, presenting a maximum of 8 for Na cations and a minimum of 3.8 for Li cations. 

\subsubsection{Local structure of oxygen}
\begin{figure*}[htb!]
\includegraphics[width=\textwidth]{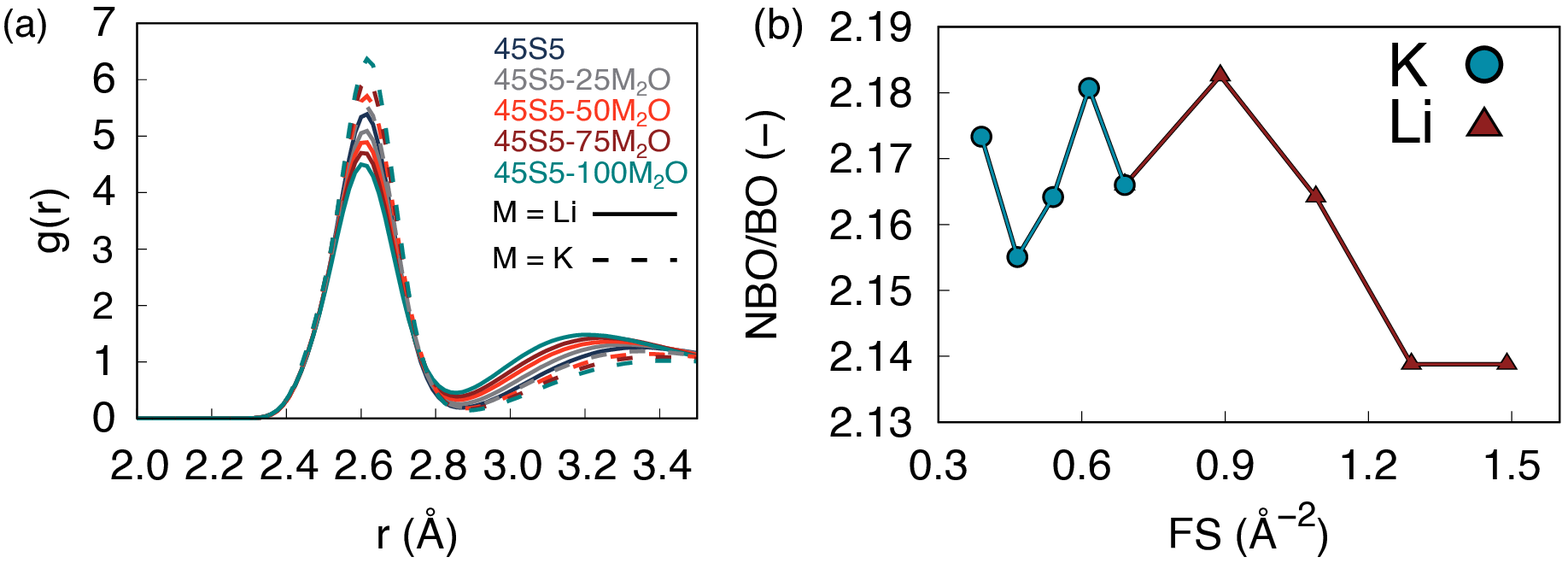}
\caption{\label{RDFOO}(a) The O--O pair distribution functions in Li mixed-alkali (solid lines) and K mixed-alkali 45S5 (dashed lines) bioactive glasses at 300 K, (b) the NBO/BO ratio as a function of the mean field strength.}
\end{figure*}

The PDFs of the O--O pair are depicted in Fig.~\ref{RDFOO}(a), showing the mean distance between two oxygen atoms as given by the first peak position. With the decrease of the field strength, the O--O distance increases as shown by the shift of the PDF toward larger distances, and it becomes slightly broader, indicating that longer O--O bonds are present in the systems when the mean FS is low. This highlights the presence of large polyhedral units in the glasses with low mean FS. The oxygen local environment is further characterized by how many glass former (Si or P) neighbours it have in its first coordination shell. This results in defining two types of oxygen in these glasses: non-bridging oxygen (NBO), an oxygen bonded to only one Si or P, and bridging oxygen (BO), an oxygen bonded to two Si or P.
Moreover, it is worth stressing that P--O--P bonds do not exist in the simulated glasses. As shown in Fig.~\ref{RDFOO}(b), the population of NBO is higher than that of BO, which was expected as the glasses contain almost 50 mol\% of modifiers and should have more NBO than BO. These findings are in good agreement with experimental data \cite{Bhaskar2020, Brckner2016}. Moreover, the content of NBO and BO is less affected by the type of modifiers used to substitute Na and can be neglected.

\subsubsection{Network connectivity and clustering of the modifiers around the glass formers}
\begin{figure*}[ht!]
\includegraphics[width=\textwidth]{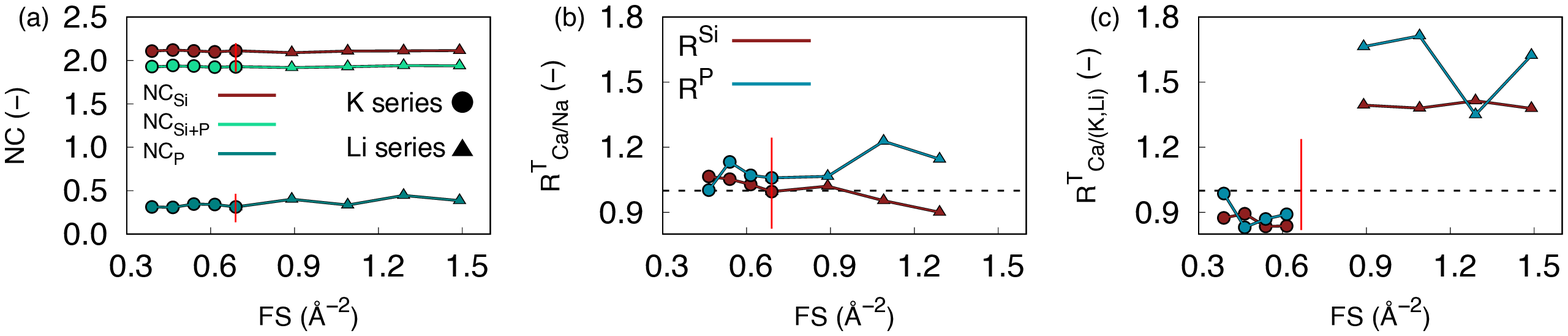}
\caption{\label{NC} (a) Total and partial network connectivity of the mixed-alkali 45S5 bioactive glass. (b) The preferential clustering ratio of the Ca and Na around Si or P. (c) The preferential clustering ratio of the Ca and Li or Ca and K around Si or P. The red vertical lines indicate the original 45S5 bioactive glass, while the horizontal dashed lines in (b) and (c) highlight R = 1 which indicate a statistical distribution of the modifiers around the glass formers. Filled circles indicate data for potassium mixed-alkali 45S5, while filled triangles are for lithium mixed-alkali 45S5 bioactive glasses.  The red vertical lines indicate the pristine 45S5 bioactive glass. The error is smaller than the symbol size.}
\end{figure*}

The network connectivity of the pristine and mixed-alkali glasses was calculated based on the Q$^n$ distributions using Eq.~\ref{NCeq} 
\begin{equation}
\label{NCeq}
NC = \sum_{n=0}^n n x_n
\end{equation}
with x$_n$ is the fraction of the Q$^n$, with n = 0, 1, 2, 3, or 4, the number of bridging oxygen (BO) atoms bound to a network-forming cation. Figure~\ref{NC}(a) shows the partial (Si, P) and total NC. The total NC is around 1.9 and is not dependent on the composition, which is in excellent agreement with the experimental data \cite{Du2012, Brckner2016}. The P based NC is around 0.4 and almost shows no dependence on the composition, and it is lower than the total NC, while that of Si is around 2.15 and is higher than the total NC. 

The medium-range structure of oxide glasses is very important in what concerns the bioactivity and also other glass properties \cite{Du2012}. Greaves, in his modified random network model \cite{Greaves1985}, suggested that in silicate glasses, it is expected to have modifier rich regions and network former rich regions leading to a heterogeneous network. The heterogeneity strongly affects oxide glasses' properties, including the diffusion, implying that all properties that depend on it will also be affected, such as bioactivity. The aggregation of a modifier around a glass former (Si or P) is quantified by calculating the ratio R$^T_{i/j}$:

\begin{equation}
R^T_{i/j} = \frac{CN_{T-i}}{CN_{T-j}} \times \frac{N_j}{N_i},
\end{equation}
where T is the network former, Si or P, i and j are two different modifier cations among Li, Na, K or Ca; N$_i$ is the number of i type of cation in the simulation box. A ratio R near to one indicates that i and j ions are distributed around T statistically without any preference. However, a ratio R $>$ 1 suggests that T prefer to be surrounded by i, and finally, a ratio R $<$ 1 indicates the affinity of j. Moreover, by using this parameter, we eliminate any compositional dependence due to a different number of atoms on the preference of one modifier to be around another. As shown in Fig.~\ref{NC}(b), for all compositions, P prefers to be surrounded by Ca than by Na. Silicon prefers Ca in the systems that contain potassium, prefer Na in the glasses that have lithium and does not show any preference in the pristine 45S5. In addition to that, Fig.~\ref{NC}(c) shows that both Si and P prefer to be surrounded by Ca than Li and K than Ca in the lithium and potassium mixed-alkali 45S5 bioactive glasses, respectively. These results are consistent with previous works \cite{Lu2018}.

\subsubsection{Clustering of modifiers}
\begin{figure*}[htb!]
\includegraphics[width=\textwidth]{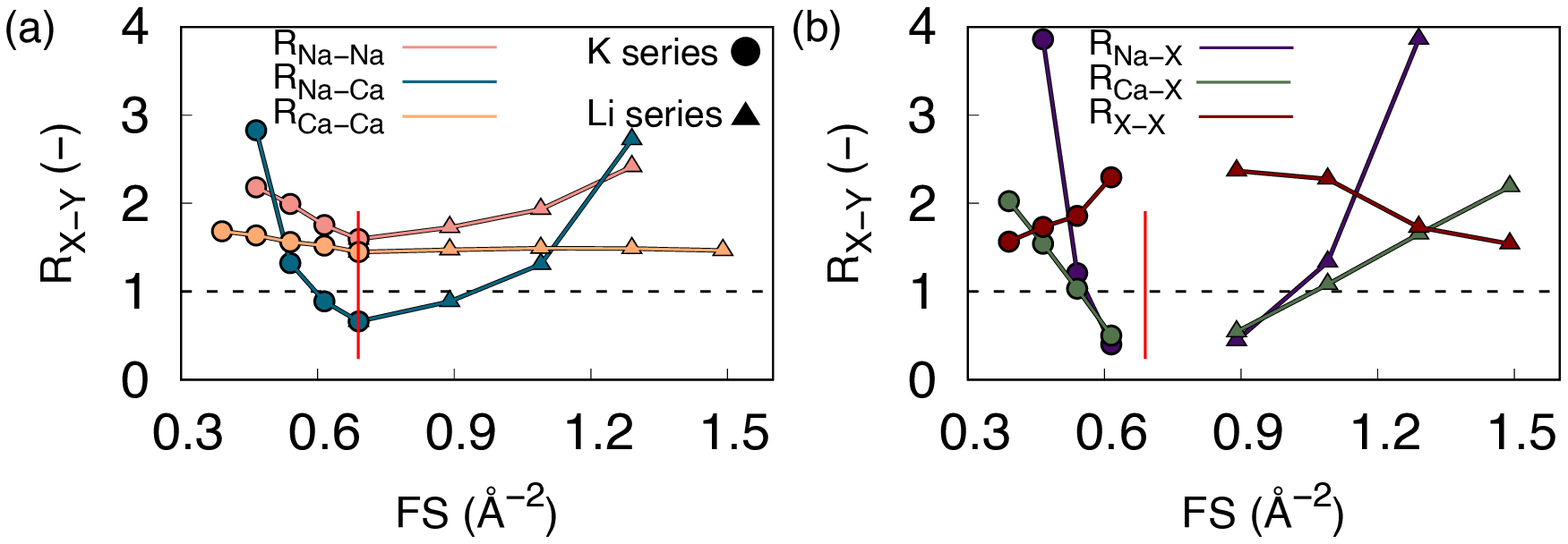}
\caption{\label{Clusteringmodifiers}The clustering ratio of Na--Na, Na--Ca, and Ca--Ca (a) and Na--X, Ca--X, and X--X (where X is Li or K) pairs as a function of the mean field strength at 300 K. Filled circles indicate data for potassium mixed-alkali 45S5, while filled triangles are for lithium mixed-alkali 45S5 bioactive glasses. The red vertical lines indicate the pristine 45S5 bioactive glass. The data is averaged over the last 100 ps from an NVT run, and the error bars are within the size of the symbol and are removed for clarity. R = 1 indicates no clustering, while R $>$ 1 implies spatial clustering.}
\end{figure*}

In oxide glasses, it is known that the modifiers can show spacial clustering \cite{Greaves1985, Atila2020c}, which was shown to depend on the modifier content and type \cite{Atila2019b}. This type of clustering can be extracted from trajectories obtained by MD simulations. To do so, we use the parameter proposed by Tilocca \textit{et al.} \cite{Tilocca2007clus} which is defined in eq.~\ref{clusteringeq}:

\begin{equation}
\label{clusteringeq}
    R_{X-Y} = \frac{N_{X-Y,MD}}{N_{X-Y,hom}} = \frac{CN_{X-Y}+\delta_{(X-Y)}}{\frac{4}{3}\pi r^3_c \frac{N_X}{V_{box}}},
\end{equation}

where $\delta_{(X-Y)}$ is 1 if $X = Y$ and 0 otherwise and r$_c$ is a cutoff distance at which the $X-Y$ coordination number CN$_{X-Y}$ of is calculated. The N$_X$ represent the total number of atoms $X$ contained in the simulation box of volume V$_{box}$. In Fig.~\ref{Clusteringmodifiers} (a) we plotted this ratio to check for Na--Na, Na--Ca, and Ca--Ca clustering tendency as a function of the third modifier content that is Li, or K. In Fig.~\ref{Clusteringmodifiers} (b), we plotted the Na--X, Ca--X, and X--X clustering as a function of the third modifier content and with X being Li, or K.

\begin{figure}[htb!]
\includegraphics[width=\columnwidth]{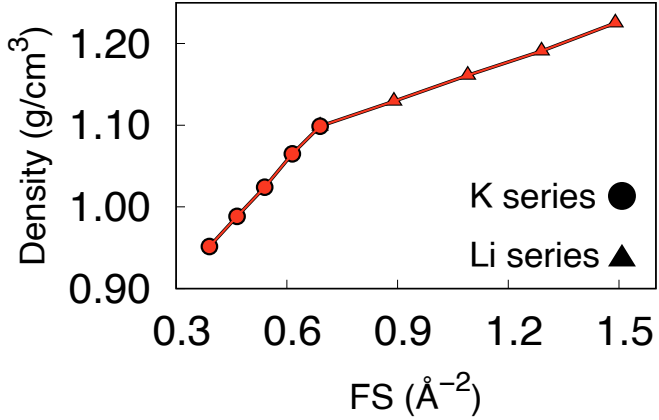}
\caption{\label{Oxygendensity}Oxygen density as a function of the mean field strength at 300 K. Filled circles indicate data for potassium mixed-alkali 45S5, while filled triangles are for lithium mixed-alkali 45S5 bioactive glasses. The data is averaged over the last 100 ps from an NVT run, and the error bars are within the size of the symbol and are removed for clarity.}
\end{figure}

\begin{figure}[htb!]
\includegraphics[width=\columnwidth]{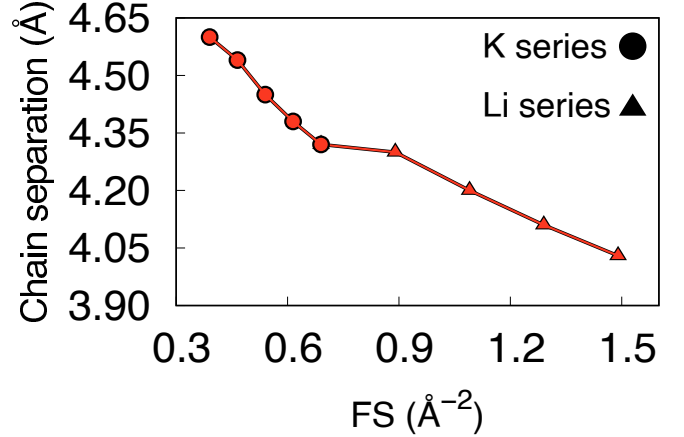}
\caption{\label{Separationdistance}Chain separation as a function of the mean field strength for the simulated glasses at 300 K. The chain separation distance was calculated from the radial and angular distribution functions. Filled circles indicate data for potassium mixed-alkali 45S5, while filled triangles are for lithium mixed-alkali 45S5 bioactive glasses.
}
\end{figure}

\begin{figure}[htb!]
\includegraphics[width=\columnwidth]{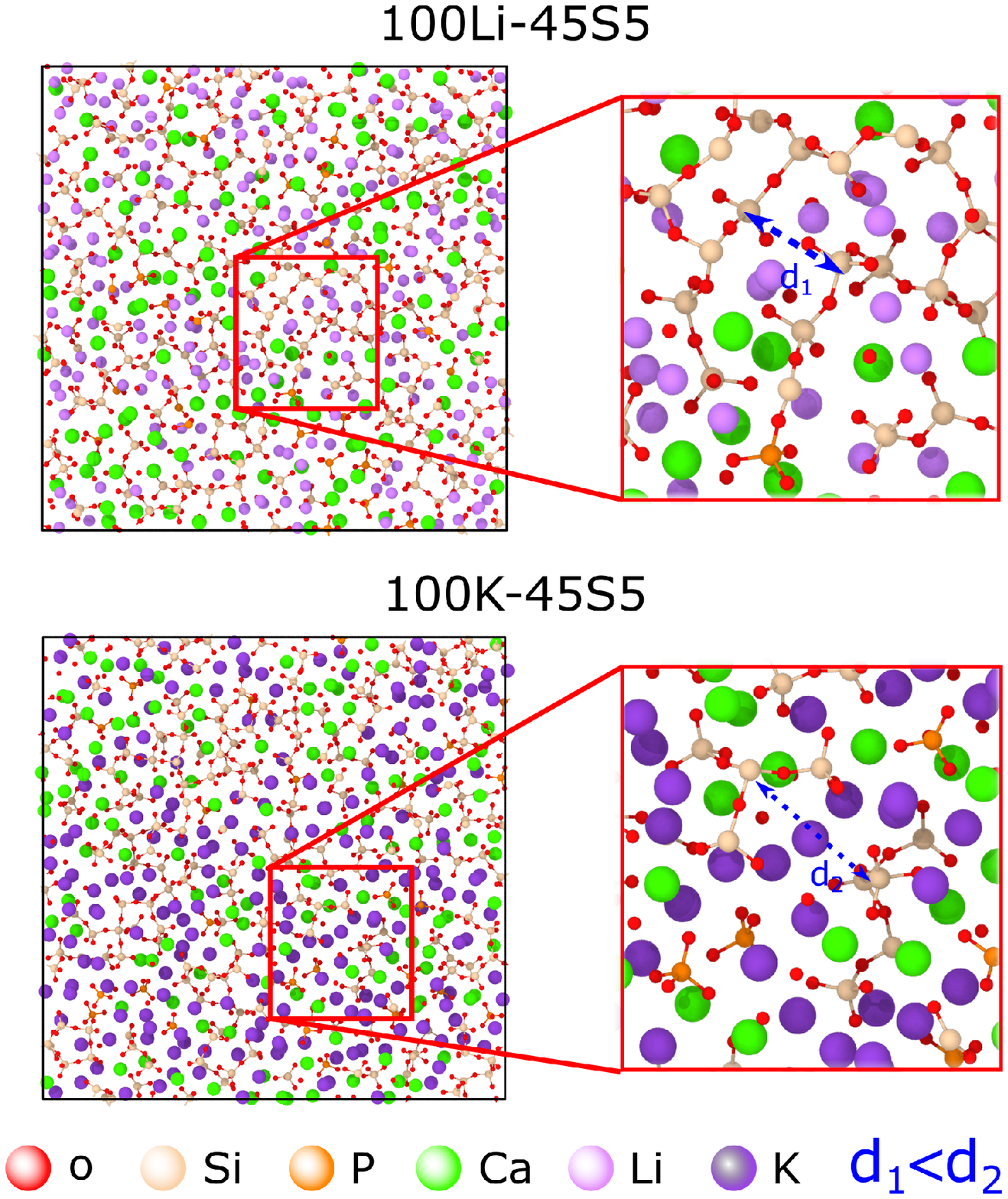}
\caption{\label{Snapshots}Snapshots showing two slices of a width of 6 \AA\, and zoomed in images showing two chains and their separation distance in the sample containing 100 mol\% Li and the samples containing 100 mol\% K. The blue dashed arrowed lines highlight the separation distance between the two selected chains with d$_1$ and d$_2$ the distances in 100Li-45S5 and 100K-45S5, respectively. the snapshots clearly show that the chains are closer to each other in 100Li-45S5 glass, and they are pushed away in the 100K-45S5 glass. }
\end{figure}

\section{\label{discussion}Discussion}
The MAE we are focusing on in this paper is studied by partial substitution of Na in the 45S5 bioactive glass by either Li or K. By comparing the radii and molar masses of Li, Na, and K, it is clear and well known that Li have the smallest radius and lowest mass followed by Na and then K. Based on this, it is expected that the density of the glasses should increase with increasing K content and decrease with increasing Li content. However, if we look at Tab.~S1 in supplementary material \cite{SuppMat}, one can see that the density is decreasing in both Li- and K-mixed alkali 45S5, which is not the expected behaviour. Moreover, glasses with a similar amount of alkali substitution have close enough densities, which was found to be independent of the type of the alkali metal (Li or K). Many factors could explain this. For instance, the change in the Li--Li and K--K clustering ratios supports this density decrease with increasing Li or K content. This ratio decreases with increasing Li or K content indicating an increased repulsion between these modifiers, which will lead to an increase in the volume and thus decrease of the density. In addition, the oxygen density (shown in Fig.~\ref{Oxygendensity}), which indicates the compactness of the glasses, show that while the density decreases in both glass series, the compactness of these glasses is entirely different. The oxygen density increases in the case of Li mixed alkali 45S5 glasses while it decreases in the K mixed alkali 45S5 glasses, which is in good agreement with experimental observations \cite{Tylkowski2013}. As a matter of fact, Li has a very small ionic radius compared to K; thus, when Li substitutes Na the SiO$_4$ chains get closer to each other.
In contrast, in the case of substituting Na by K, the SiO$_ 4$ chains get pushed away from each other because of the larger size of K, which is clearly shown in Fig.~\ref{Separationdistance}, where the chains separation distance decreases with increasing Li content and increases with increasing K content. This will make the glass network less compact in the case of K substituted 45S5 as highlighted by the change of the oxygen packing density \cite{Tylkowski2013, Inaba2020} and vice-versa for Li substituted 45S5. The snapshots in Fig.~\ref{Snapshots} visually show that the two selected chains get closer to each other in the 100Li-45S5 glass, and they are pushed away in the 100K-45S5 glass. Thus, this competition between the atomic weight and size of cation, which one can be dominant over the other, plays an important role in affecting the glass density and other properties, as will be discussed later. 
The presence of a silicate network made mainly by Q$^2$ units and orthophosphate units is usually due to modifier cations which balance the charge \cite{Brckner2016}. For the 45S5, when we only have sodium and calcium as modifiers, we showed that Ca and Na are randomly distributed around Si and P (see Fig.\ref{NC}b). The presence of a third cation with a significantly different radius indeed cause a preference, as was evidenced in Fig.~\ref{NC}(b and c). The preference of P to be surrounded by Ca is more pronounced in the Li series than in the K series, which is related to the compactness of the Li substituted 45S5 glass.

The properties of the glasses that depend on the transport mechanisms, such as the ion release and glass transition, strongly depend on the composition of the glass. The partial substitution of an alkali by another one, in some cases, causes a non-linear change in the properties, where the glass properties exhibit minima or maxima at the composition with equal content of the alkali metals. However, previous experimental studies \cite{Brckner2016} showed that mixed alkali 45S5 did not show a non-linear dependence in their properties, meaning that no maxima or minima were observed when compared to single alkali glasses. This indeed correlates pretty well with our findings, where no minima or maxima in the glass properties were observed compared to the pristine 45S5. Instead, the diffusion decreased linearly with increasing mean FS, and the elastic moduli increased with increasing mean FS. The decrease in the diffusion in the 100Li-45S5 glass is due to the compact network of the glass, making the ion mobility harder.
On the other hand, the 100K-45S5 glass has a more open network, leading to a slight increase in ion mobility and a decrease in the K self-diffusion activation energy.  The modifiers in oxide glasses are surrounded by oxygen polyhedra, the number of the oxygen surrounding the modifiers depends on the size of the modifier and thus on the field strength \cite{Atila2019b}. This will eventually lead to the formation of microsegregation of the modifiers forming channels within the glass network \cite{Greaves1985}. The tendency of channel formation will indeed depend on the type of the modifier and will be affected by the two or more modifiers present within the glass network \cite{Tylkowski2013, Du2012}. The modifiers clustering tendency has been previously discussed in several papers \cite{Atila2020c, Wang2019}, and it is evidenced in this study as shown in Fig.~\ref{Clusteringmodifiers}. It was shown that the modifiers channels play the role of energetically favourable routes for ionic diffusion \cite{Tylkowski2013, Mahadevan2019}. In oxide glasses, the process of ionic diffusion is composed of a succession of ion hops along the channels, where cation hop from one polyhedron to a recently vacated neighbouring polyhedron \cite{Atila2020a}. The differences in the electronegativity of the alkali metals can also affect the ion release rate, which was observed by the behaviour of the diffusion coefficients and the activation energies. The alkali elements with a higher degree of covalency in their bonding (e.g., Li), as depicted by the values of FS, tend to be strongly bonded to oxygen atoms and, therefore, have a reduced ion release compared to elements with lower bond covalency (lower FS) such as K. The absence of MAE in the 45S5 (Hench type) glasses can be mainly attributed to the fact that the network is less polymerized compared to that in network silicate glasses which has higher network connectivity and thus different transport mechanisms \cite{Mascaraque2017}. This, will lead to significant differences in the dissolution behaviour and ion release \cite{Mascaraque2017, Wang2017b} which were shown to depend on the glass topology \ strongly cite{Mascaraque2017}. Although the network connectivity of these glasses is almost the same, the diffusion coefficient and activation energies for self-diffusion were strongly affected by the sodium substitution by Li and K, and we observed that the pristine 45S5 showed the optimal diffusion properties that will eventually lead to optimal ion-release and thus apatite formation when compared to the substituted glasses. This effect is more pronounced in the case of Li-substituted 45S5 glasses. Hence, based on the above discussion, it is evident that the mixed alkali glasses from both series have lower bioactivity compared to the pristine 45S5. However, the K-45S5 series is expected to be more bioactive than the Li-45S5 series. These findings are in good agreement with experimental observations, where the pristine 45S5 glass showed the fastest apatite formation and best bioactivity when compared to the substituted glasses \cite{Brckner2016}. However, it is worth noting that the full or partial substitution of sodium or calcium can still be beneficial and desirable in tailoring the mechanical properties of these glasses, especially the hardness and elastic moduli to make them mechanically compatible with the implantation zone. In addition to tuning the mechanical properties of the glasses, the partial substitution can be used to prevent crystallization or control some of the body functions that will help in fast healing \cite{Tabbassum2021}.

The change in the glass structure can also be related to the change of the elastic moduli. The Li-45S5 glass showed the highest elastic moduli among other glasses, while K-45S5 showed the lowest elastic moduli. This correlates well with the glass packing density as shown by the oxygen content, where glasses with higher oxygen density have the highest elastic moduli and vice-versa. These trends are in good agreement with experimental findings \cite{Tylkowski2013}. Furthermore, the glass strength, as indicated by the F$_{net}$, increases with increasing Li content while it decreases with increasing K content; this is also observed in the change of FS, indicating an increase of the bonding strength that eventually leads to an increase of the elastic moduli. 

\section{\label{conclusion}Conclusion}
In conclusion, using MD simulations, we showed that there is no visible mixed alkali effect on the diffusion and elastic properties of Li- and K-substituted 45S5 bioactive glasses. However, we pointed out that the complete or partial substitution of sodium by another alkali will be desired for tuning the properties of these bioactive glasses, such as ion release and mechanical compatibility. 
We demonstrated that the changes in the diffusion and elastic properties were related to the changes of the packing density of these glasses, which is positively correlated with the elastic moduli. The changes of the oxygen packing density were mainly due to the --Si--O--Si-- chains getting closer in the glasses having Li and pushed away from each other in the glasses containing K. Furthermore, the changes in the dynamic behaviour of the studied systems, will indeed, affect the processing window of these glasses, e.g., reducing crystallization.
We believe that the insights provided in this paper into the alkali substituted 45S5 will pave routes for tailoring the properties of bioactive glass and for the rational design of new bioactive glasses for medical and clinical applications.

\begin{acknowledgments}
A. Atila thank the German Research Foundation (DFG) for financial support through the priority program SPP 1594 -- Topological Engineering of Ultra-Strong Glasses. The authors gratefully acknowledge the computing resources provided by the Erlangen Regional Computing Center (RRZE) to run some of the simulations.
\end{acknowledgments}

\bibliography{references}

\end{document}